\documentclass[aps,showpacs,showkeys]{revtex4}
\usepackage{amsmath,amsthm,amssymb,epsfig,alltt}

\begin{document}

\def\a{\alpha}
\def\b{\beta}
\def\d{{\delta}}
\def\l{\lambda}
\def\e{\epsilon}
\def\p{\partial}
\def\m{\mu}
\def\n{\nu}
\def\t{\tau}
\def\th{\theta}
\def\s{\sigma}
\def\g{\gamma}
\def\o{\omega}
\def\r{\rho}
\def\z{\zeta}
\def\D{\Delta}
\def\half{\frac{1}{2}}
\def\hatt{{\hat t}}
\def\hatx{{\hat x}}
\def\hatp{{\hat p}}
\def\hatX{{\hat X}}
\def\hatY{{\hat Y}}
\def\hatP{{\hat P}}
\def\haty{{\hat y}}
\def\whatX{{\widehat{X}}}
\def\whata{{\widehat{\alpha}}}
\def\whatb{{\widehat{\beta}}}
\def\whatV{{\widehat{V}}}
\def\hatth{{\hat \theta}}
\def\hatta{{\hat \tau}}
\def\hatrh{{\hat \rho}}
\def\hatva{{\hat \varphi}}
\def\barx{{\bar x}}
\def\bary{{\bar y}}
\def\barz{{\bar z}}
\def\baro{{\bar \omega}}
\def\barpsi{{\bar \psi}}
\def\sp{\sigma^\prime}
\def\nn{\nonumber}
\def\cb{{\cal B}}
\def\2pap{2\pi\alpha^\prime}
\def\wideA{\widehat{A}}
\def\wideF{\widehat{F}}
\def\beq{\begin{eqnarray}}
 \def\eeq{\end{eqnarray}}
 \def\4pap{4\pi\a^\prime}
 \def\op{\omega^\prime}
 \def\xp{{x^\prime}}
 \def\sp{{\s^\prime}}
 \def\ap{{\a^\prime}}
 \def\tp{{\t^\prime}}
 \def\zp{{z^\prime}}
 \def\xpp{x^{\prime\prime}}
 \def\xppp{x^{\prime\prime\prime}}
 \def\barxp{{\bar x}^\prime}
 \def\barxpp{{\bar x}^{\prime\prime}}
 \def\barxppp{{\bar x}^{\prime\prime\prime}}
 \def\zetap{{\zeta^\prime}}
 \def\barchi{{\bar \chi}}
 \def\baro{{\bar \omega}}
 \def\bpsi{{\bar \psi}}
 \def\barg{{\bar g}}
 \def\barz{{\bar z}}
 \def\bareta{{\bar \eta}}
 \def\ta{{\tilde \a}}
 \def\tb{{\tilde \b}}
 \def\tc{{\tilde c}}
 \def\tz{{\tilde z}}
 \def\tJ{{\tilde J}}
 \def\tpsi{\tilde{\psi}}
 \def\tal{{\tilde \alpha}}
 \def\tbe{{\tilde \beta}}
 \def\tga{{\tilde \gamma}}
 \def\tchi{{\tilde{\chi}}}
 \def\barth{{\bar \theta}}
 \def\bareta{{\bar \eta}}
 \def\barom{{\bar \omega}}
 \def\bole{{\boldsymbol \epsilon}}
 \def\bolth{{\boldsymbol \theta}}
 \def\bomega{{\boldsymbol \omega}}
 \def\bolmu{{\boldsymbol \mu}}
 \def\bola{{\boldsymbol \alpha}}
 \def\bolb{{\boldsymbol \beta}}
 \def\bolX{{\boldsymbol X}}
 \def\boln{{\boldsymbol n}}
 \def\bba{{\boldsymbol a}}
 \def\bbA{{\boldsymbol A}}

\setcounter{page}{1}
\title[]{String Scattering Amplitudes and Deformed Cubic String Field Theory
}

\author{Sheng-Hong Lai}
\email{xgcj944137@gmail.com}
\affiliation{Department of Electrophysics, National Chiao-Tung University, Hsinchu, Taiwan, R.O.C.}
\author{Jen-Chi Lee}
\email{jcclee@cc.nctu.edu.tw}
\affiliation{Department of Electrophysics, National Chiao-Tung University, Hsinchu, Taiwan, R.O.C.}
\author{Taejin Lee\footnote{Corresponding author}}
\email{taejin@kangwon.ac.kr}
\affiliation{Department of Physics, Kangwon National University, Chuncheon 200-701 Korea}
\author{Yi Yang}
\email{yiyang@mail.nctu.edu.tw}
\affiliation{Department of Electrophysics, National Chiao-Tung University, Hsinchu, Taiwan, R.O.C.}
\author{}
\date{\today }

\begin{abstract}
We study string scattering amplitudes by using the deformed cubic string field theory
which is equivalent to the string field theory in the proper-time gauge. The four-string scattering amplitudes
with three tachyons and an arbitrary string state are calculated. The string field theory yields the 
string scattering amplitudes evaluated on the world sheet of string scattering whereas the coventional 
method, based on the first quantized theory brings us the string scattering amplitudes defined on 
the upper half plane. For the highest spin states, generated by the primary operators, both calculations 
are in perfect agreement. In this case, the string scattering amplitudes are invariant under the conformal transformation, which maps the string world sheet onto the upper half plane.
If the external string states are general massive states, generated by non-primary field operators,
we need to take into account carefully the conformal transformation between the world sheet and the upper 
half plane. We show by an explicit calculation that the string scattering amplitudes calculated by using 
the deformed cubic string field theory transform into those of the first quantized theory 
on the upper half plane by the conformal transformation, generated by the Schwarz-Christoffel mapping.

\end{abstract}




\maketitle

\section{Introduction}

String theory was formulated first in terms of the scattering amplitude, which is now known as Veneziano amplitude \cite{Veneziano68}. The Veneziano amplitude was soon generalized to the Virasoro-Shapiro amplitude
\cite{Virasoro69, Shapiro69}. These scattering amplitudes marked the birth of the string theory. The Veneziano amplitude corresponds to the four-tachyon scattering amplitude of open string, and the Virasoro-Shapiro scattering amplitude 
corresponds to the four-tachyon scattering amplitude of closed string. 
Since the birth of string theory, the string scattering amplitude has played a
major role in the development of string theory. The string scattering amplitudes usually have been studied by 
using the first quantized string theory. In the first quantized theory, we may obtain the string scattering amplitudes by evaluating the Polyakov string path integral with vertex operators inserted on the string world sheet. For open string theory, the upper half plane is chosen as the string world sheet and the vertex operators, representing external string states, are inserted on the real axis. The string scattering amplitudes
are found to enjoy various relationships between themselves, which are important in understanding the profound 
nature of string theory. Among them are string Bern-Carrasco-Johansson (BCJ) relations \cite{Chan2006notes,Bjerrum2009,Stieberger09,Lai2016string}, which are the string theory generalizations of gauge field theory BCJ relations \cite{BCJ} and the Kawai-Lewellen-Tye (KLT) relations \cite{Kawai86}, which relate the tree level closed string scattering amplitudes to those of open string.
The string scattering amplitudes also satisfy the generalized on-shell Ward identities \cite{JCLee1994,Chan2004zero,Chan2005stringy} which, together with the string BCJ relation, 
drastically reduce the number of independent string scattering amplitudes.

In the present work, we shall study the string scattering amplitudes in the framework of the covariant interacting string field theory. In particular, by using the deformed cubic string field theory \cite{TLee2017plb},
we will calculate the four-string scattering amplitudes with three tachyons and an arbitrary string states, which have been extensively studied in Refs. 
\cite{Chan2005solving,Chan2006comm,Chan2006high,JCLee2014appel,JCLee2015rev,Lai2016lauri,Lai2016exact,
Lai2017sloving} to explore the symmetric properties of the string scattering amplitudes in the high energy limit \cite{Gross87,Gross88,Gross88prl,Gross89Phil,Gross89nucl}.
In recent works \cite{TLee2017plb,TLee1609}, one of the authors showed that the non-planar world sheets of Witten's cubic open string field theory \cite{Witten1986,Witten92p} can be made planar if we choose the external string states judiciously. The deformed cubic string 
field theory, which is equivalent to the covariant string field theory in the proper-time gauge \cite{TLee88ann}, has a 
number of advantages over other approaches: 1) The world sheet diagrams of the string scattering are planar so that we can apply the light-cone string field theory techniques 
\cite{Mandelstam1973,Mandelstam1974,Kaku1974a,Kaku1974b,Cremmer74,Cremmer75,GreenSW}. 2) The theory does not contain unphysical length parameters \cite{Hata1986,Hata86prd} and yet possesses the BRST gauge symmetry because its action is formally equivalent to that of Witten's cubic string field theory \cite{Witten1986}. 3) We can obtain the exact gauge invariant Yang-Mills field action without using the level truncations \cite{Coletti03} or the field redefinitions \cite{Feng2007} in the zero-slope limit. 
We expect that the deformed cubic string field theory or the 
covariant string field theory in the proper-time gauge produces the string scattering amplitudes which 
can be directly compared with those of the first quantized string theory. 

\section{String Scattering Amplitudes}

We shall begin with a brief review of the recent work \cite{Lai2016lauri}
on the string scattering amplitudes (SSA) of three 
tachyons and one arbitrary string states which are based on the first quantized string theory. 
It fixes the notations and the kinematics of the scattering in the center of mass frame. 
In the center of mass frame, the kinematics are defined as
\begin{subequations}
\begin{align}
k_{1}  &  =\left(  \sqrt{M_{1}^{2}+|\vec{k_{1}}|^{2}},-|\vec{k_{1}}|,0\right)
,\\
k_{2}  &  =\left(  \sqrt{M_{2}+|\vec{k_{1}}|^{2}},+|\vec{k_{1}}|,0\right)  ,\\
k_{3}  &  =\left(  -\sqrt{M_{3}^{2}+|\vec{k_{3}}|^{2}},-|\vec{k_{3}}|\cos
\phi,-|\vec{k_{3}}|\sin\phi\right)  ,\\
k_{4}  &  =\left(  -\sqrt{M_{4}^{2}+|\vec{k_{3}}|^{2}},+|\vec{k_{3}}|\cos
\phi,+|\vec{k_{3}}|\sin\phi\right)
\end{align}
\end{subequations}
with $M_{1}^{2}=M_{3}^{2}=M_{4}^{2}=-2$ and $\phi$ is the scattering angle.
The Mandelstam variables are defined as 
\beq
s=-\left(  k_{1}+k_{2}\right)^{2},~~~
t=-\left(  k_{2}+k_{3}\right)^{2}, ~~~ u=-\left(  k_{1}+k_{3}\right)^{2}.
\eeq
We choose the three polarization vectors on the scattering plane as follows:
\begin{subequations}
\begin{align}
e^{T}  &  =(0,0,1),\\
e^{L}  &  =\frac{1}{M_{2}}\left(  |\vec{k_{1}}|,\sqrt{M_{2}+|\vec{k_{1}}|^{2}%
},0\right)  ,\\
e^{P}  &  =\frac{1}{M_{2}}\left(  \sqrt{M_{2}+|\vec{k_{1}}|^{2}},|\vec{k_{1}%
}|,0\right)  .
\end{align}
\end{subequations}
Note that SSA of three tachyons and one arbitrary string state with
polarizations orthogonal to the scattering plane vanish.
For later use, we define
\begin{equation} \label{polarization}
k_{i}^{X}\equiv e^{X}\cdot k_{i}\text{ \ for \ }X=\left(  T,P,L\right)  .
\end{equation}

The general string states at mass level%
\begin{equation}
M_{2}^{2}=2(N-1),~~~N=\sum_{n,m,l>0}\left(  nr_{n}^{T}+mr_{m}^{P}+lr_{l}%
^{L}\right)  ,
\end{equation}
with polarizations on the scattering plane are of the form%
\begin{equation}
\left\vert r_{n}^{T},r_{m}^{P},r_{l}^{L}\right\rangle =\prod_{n>0}
\left(\alpha_{-n}^{T}\right)^{r_{n}^{T}}\prod_{m>0}\left( \alpha_{-m}^{P}\right)
^{r_{m}^{P}}\prod_{l>0}\left( \alpha_{-l}^{L}\right)^{r_{l}^{L}}%
|0,k\rangle. \label{state}%
\end{equation}
The four-string scattering amplitude with three tachyons and one general string state 
Eq. (\ref{state}) in the $s$-$t$ channel is found to be \cite{Lai2016lauri}
\beq \label{Ast1st}
A_{st}^{\left(  r_{n}^{T};r_{m}^{P};r_{l}^{L}\right)} &=& 
\int_{0}^{1}dx\,x^{k_{1}\cdot k_{2}}(1-x)^{k_{2}\cdot k_{3}}%
\prod_{n=1}  \left\{ \left(-1\right)^{n-1} (n-1)! \left( \frac{k_{1}^{T}}{x^{n}%
}+\frac{k_{3}^{T}}{(x-1)^{n}} \right)\right\}^{r_{n}^{T}}\nonumber\\
&& \prod_{m=1}\left\{\left(-1\right)^{m-1}(m-1)!\left( \frac{k_{1}^{P}%
}{x^{m}}+\frac{k_{3}^{P}}{(x-1)^{m}} \right)\right\}  ^{r_{m}^{P}%
} \prod_{l=1}\left\{ \left(-1\right)^{l-1}(l-1)!\left( \frac{k_{1}^{L}%
}{x^{l}}+\frac{k_{3}^{L}}{(x-1)^{l}} \right)\right\}  ^{r_{l}^{L}%
}.
\eeq
Here the four points on the real line where the vertex operators are inserted, are chosen as 
\beq
Z_1 = 0, ~~~ Z_2 = x, ~~~ Z_3 = 1, ~~~ Z_4 = \infty . 
\eeq 
In the $s$-$t$ channel, the Koba-Nielsen variable $x$ is in the range of  $[0,1]$ while in the 
$t$-$u$ channel, $x$ is in the range of $[1, \infty)$. 
The four-string scattering amplitude is related to that of $s$-$t$ channel through the string
BCJ relation:
\beq
A_{st}^{\left(  r_{n}^{T};r_{m}^{P};r_{l}^{L}\right)  } &=&\frac{\sin(\pi
k_{2}\cdot k_{4})}{\sin(\pi k_{1}\cdot k_{2})}A_{tu}^{\left(  r_{n}^{T}
;r_{m}^{P};r_{l}^{L}\right)  } \nn\\
&=&\frac{\sin(\frac{u}{2}+2-N)\pi}{\sin(\frac
{s}{2}+2-N)\pi}A_{tu}^{\left(  r_{n}^{T};r_{m}^{P};r_{l}^{L}\right)}.
\eeq 
We can prove this string BCJ relation by an explicit calculation \cite{Lai2016lauri}, rewriting the SSA in the 
$s$-$t$ channel in terms of the $D$-type Lauricella function $F^{(K)}_D$ which is one the four
extensions of the Gauss hypergeometric function. The exact calculation of SSA in terms of the Lauricella
function was useful to redrive the symmetric relations amlong SSA in various limits: These include
the linear relations in the hard scattering limit \cite{Chan2004zero,Chan2005stringy,Chan2005ward}, the recurrence relations in the Regge scattering limit 
\cite{JCLee2014appel,Ko2009patt,JCLee2013rec}, and the extended recurrence relations in the non-relativistic 
scattering limit \cite{Lai2016string}. Reader may refer to Ref. \cite{Chan2005stringy} for more details.

\section{Deformed Cubic String Field Theory}

All types of symmetric properties of string theory should also be deduced from 
the second quantized theory, i.e., the covariant interacting string field theory. The string field 
theory may provide us a more coherent framework to understand the symmetric properties of string 
theory at a deeper level. However, in practice, it has been difficult to make use of the string field 
theory to calculate the SSA. The major obstacle was the non-planarity of the world sheet diagrams of 
the cubic open string field theory. The world sheet of $N$-string vertex is a conic surface with 
an excess angle of $(N-2) \pi$. The Fock space representation of three-string vertex has been 
obtained by Gross and Jevicki \cite{Grossjevicki87a,Grossjevicki87b} by mapping the world sheet 
of six strings onto a circular disk 
with an orbifold condition and that of the four-string vertex has been constructed by Giddings \cite{Giddings86}
by mapping the world sheet of four-string vertex onto the upper half plane with branch cuts. 
However, it is hard to make use of these constructions to calculate the SSA of four strings. 
It is so complicated to compute the Neumann functions for the four-string vertex by using the 
conformal mapping given in Ref. \cite{Giddings86}. One may obtain the four-string vertex from the cubic string 
field action by using the Fock space representation of the three-string vertex by Gross and Jevicki as 
an effective interaction. However, this procedure may involve inverting infinite dimensional matrices
and does not yield exact results. 

In a recent work \cite{TLee2017plb}, it has been shown that we can deform world sheet diagrams of the cubic string 
field theory in a consistent manner: By choosing the external string states judiciously, we may deform
the non-planar world sheet diagrams into the planar ones. The deformed cubic string field theory is 
formally equivalent to the string field theory in the proper-time gauge \cite{TLee88ann,TLee1609}. The deformation then drastically simplifies 
the calculations of SSA because well-developed light-cone string field theory techniques are 
readily applicable. The zero-slope limits of the cubic string field theory on multiple $D$-branes 
have been studied in Refs. \cite{TLee2017plb,TLee1609,TLee2017cov} by using the deformed string field theory. On the space filling multiple
$D$-branes the string field theory reduces to the non-Abelian Yang-Mills gauge theory and on
the multiple $Dp$-branes to the matrix-valued scalar field theory, interaction with the non-Abelian Yang-Mills gauge fields in the zero-slope limit. It should be noteworthy that the exact results are  
obtained without using the level truncations or the non-linear field redefinitions.

\begin{figure}[htbp]
   \begin {center}
    \epsfxsize=0.8\hsize

	\epsfbox{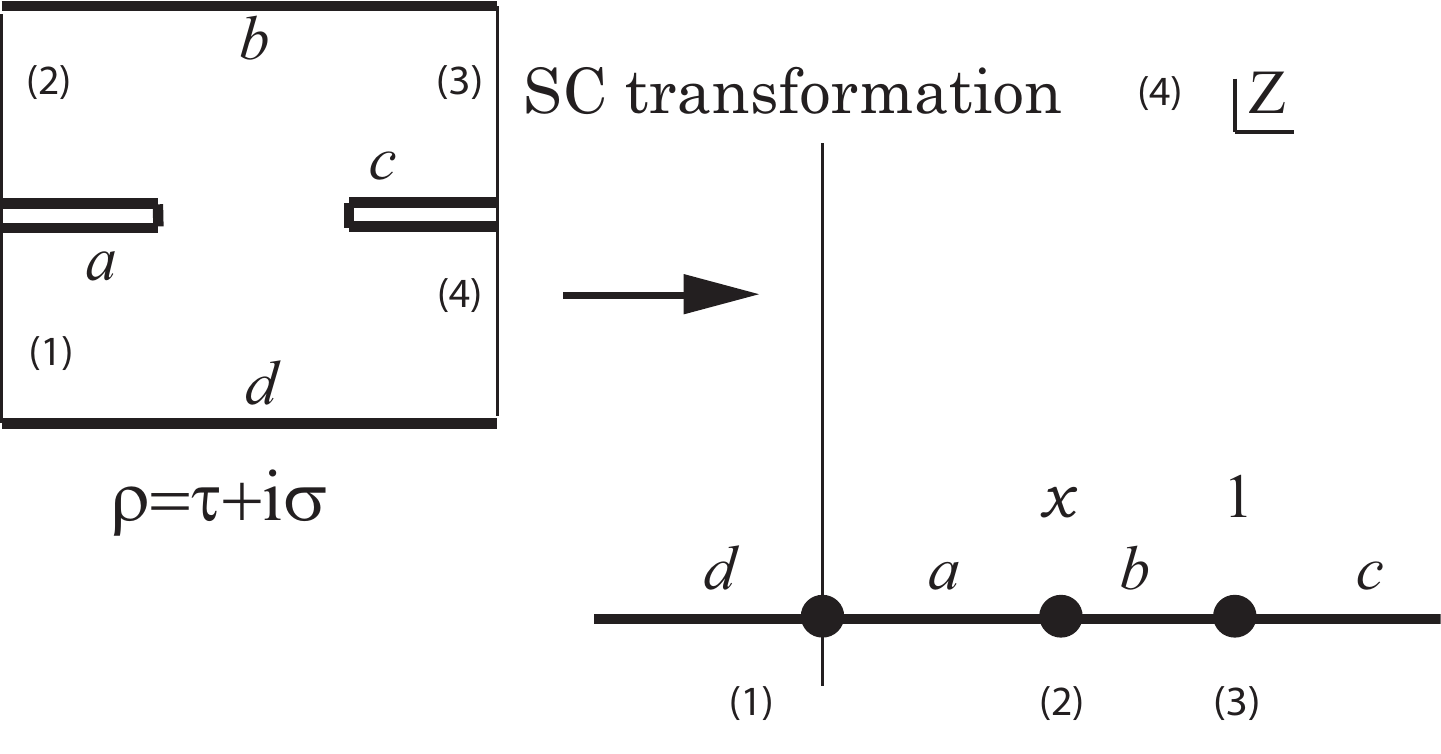}
   \end {center}
   \caption {\label{SCtrans} World sheet diagram mapped onto the upper half complex plane.}
\end{figure}

\begin{figure}[htbp]
   \begin {center}
    \epsfxsize=0.5\hsize

	\epsfbox{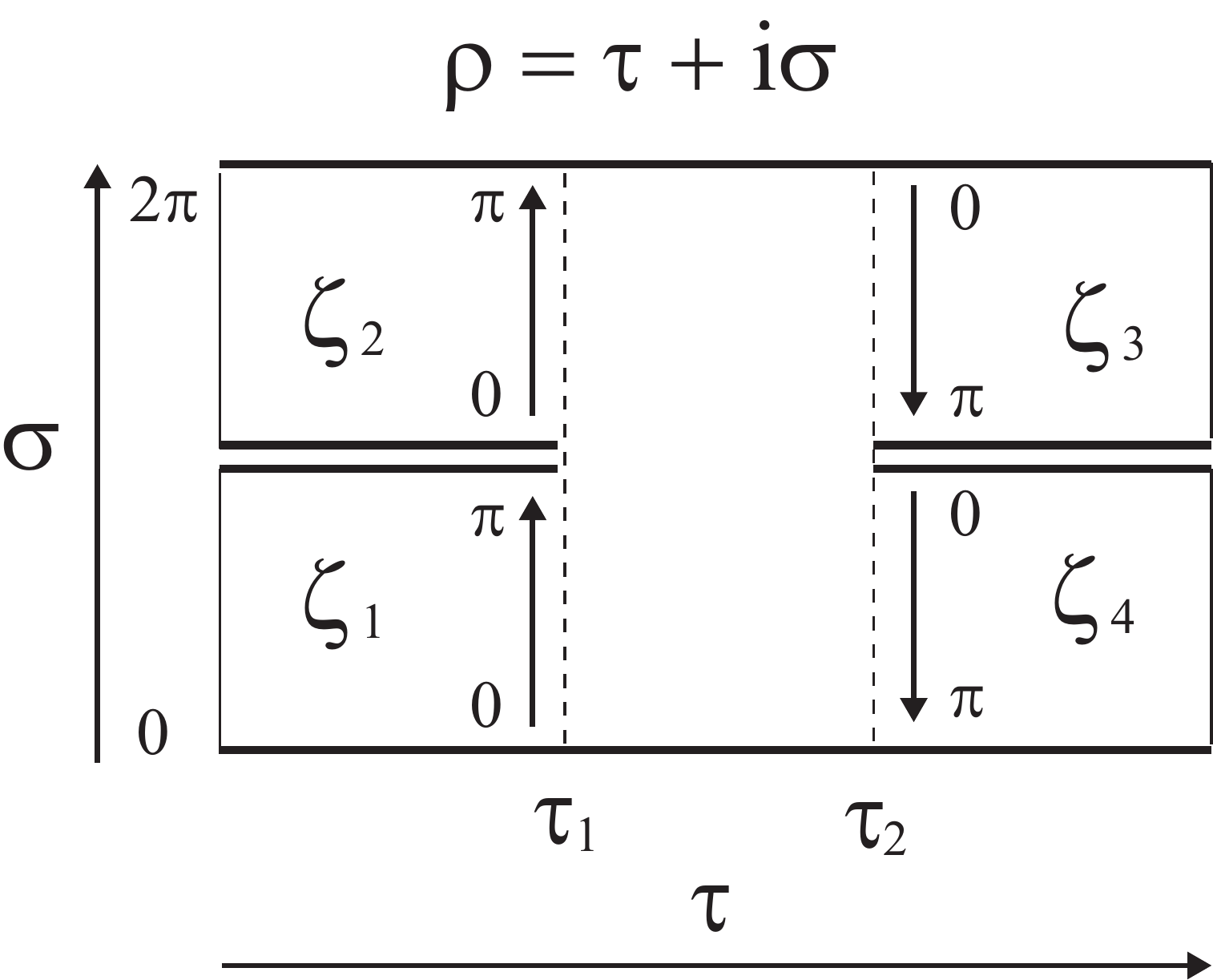}
   \end {center}
   \caption {\label{local} Local coordinates on the four-string scattering world sheet}
\end{figure}

The planar world sheet diagram of the string field theory in the proper-time
gauge is mapped onto the upper half plane by the Schwarz-Christoffel transformation which is given by 
\beq
\rho = \sum_{r=1}^4 \a_r \ln(z-Z_r) = \ln z + \ln(z-x) - \ln (z-1) + i \pi. 
\eeq
where the length parameters of the four strings are chosen as $
\a_1 = 1, ~~~ \a_2 = 1, ~~~ \a_3 =-1, ~~~ \a_4 =-1 $.
(See Fig. \ref{SCtrans}.) 
Equivalently, the Schwarz-Christoffel maping may be written as 
\beq
e^\rho = - \frac{z(z-x)}{z-1} . 
\eeq
Interaction points on the upper half plane are determined by the equation, 
\beq
\frac{\p \rho}{\p z} = \sum_{r=1}^4 \frac{\a_r}{z-Z_r} = 0 
\eeq  
which has two real solutions in the $s$-$t$ channel:
\beq
z_\pm = 1 \pm \sqrt{1-x}, ~~~ 0 \le x \le 1 .
\eeq 
These two solutions define two interaction points 
\beq
\rho (z_\pm) = \ln \frac{z_\pm(z_\pm-x)}{z_\pm-1} + i \pi 
\eeq 
and two interaction times on the world sheet, $\text{Re} \rho(z_\pm)$: 
\begin{subequations} 
\beq
\t^{(1)}_0 &=& \t^{(2)}_0 =\t_1 = 2 \ln \left(1- \sqrt{1-x} \right) < 0, \\
\t^{(3)}_0 &=& \t^{(4)}_0 = \t_2 = 2 \ln \left(1+ \sqrt{1-x} \right) > 0 . 
\eeq
\end{subequations}
We may introduce local coordinates on individual string patches, $\zeta_r$, 
$r =1, 2, 3, 4$ as depicted in Fig. \ref{local}. The conformal mapping
from the local string patches to the upper half plane may be written as follows \cite{TLee1609}:
\begin{subequations} 
\begin{align}
e^{-\zeta_1} &= - e^{\t_1} \frac{(z-1)}{z(z-x)}, &\text{on the 1st string patch},  \\
e^{-\zeta_2} &= e^{\t_1} \frac{(z-1)}{z(z-x)},   & \text{on the 2nd string patch},  \\
e^{-\zeta_3} &= - e^{-\t_2} \frac{z(z-x)}{z-1},  & \text{on the 3rd string patch}, \\
e^{-\zeta_4} &=  e^{-\t_2} \frac{z(z-x)}{z-1},   & \text{on the 4th string patch} .
\end{align}
\end{subequations}

The Neumann functions can be computed by using contour integrals on the upper half complex plane \cite{Hata1986,Mandelstam1973,Cremmer74}
\begin{subequations} 
\beq
\bar N^{rs}_{00} &=& \left\{ 
\begin{array}{ll}
\ln \vert Z_r -Z_s \vert ,  & ~~\mbox{for} ~~r\not=s,\\ 
-\sum_{i(\not=r)} 
\frac{\a_i}{\a_r} \ln \vert Z_r - Z_i \vert + \frac{1}{\a_r} \t^{(r)}_0 , & ~~
\mbox{for} ~~r=s
\end{array} \right.  \label{N00}\\
\bar N^{rs}_{n0} &=& \bar N^{sr}_{0n} = \frac{1}{n} \oint_{Z_r} \frac{d z}{2\pi i} \frac{1}{z-Z_s} e^{-n\zeta_r(z)}, ~~~n \ge 1, \label{Nn0}\\
\bar N^{rs}_{nm} &=& \frac{1}{nm} \oint_{Z_r} \frac{dz}{2\pi i} \oint_{Z_s} \frac{d \zp}{2\pi i} 
\frac{1}{(z-\zp)^2} e^{-n\zeta_r(z) - m \zeta^\prime_s(\zp)}, ~~~ n, m \ge 1  \label{Nnm}
\eeq
\end{subequations}
The deformed string field theory leads us to the four-string scattering amplitude written in terms of the Neumann functions as
\begin{subequations} 
\beq
{\cal A} &=& \int\left\vert\frac{\prod_{r=1}^4 dZ_r }{ dV_{abc}}\right\vert
\prod_{r <s} \vert Z_r -Z_s \vert^{k_r \cdot k_s + \frac{\a_s}{\a_r}\left(1-k_r^2/2\right)
+ \frac{\a_r}{\a_s}\left(1-k_s^2/2\right)}\nn\\
&& \exp \left[\sum_{r=1} \frac{\t^{(r)}}{\a_r} \left(\frac{k_r^2}{2} -1 \right) \right]
 \Bigl\langle  \Psi_1, \Psi_2, \Psi_3, \Psi_4 \Bigl\vert \exp \left[E^\prime_{[4]}\right] \Bigr\vert 0, k \Bigr\rangle, \label{4ssa1}\\ 
 E^\prime_{[4]} &=& \sum_{r, s =1}^4 \Biggl\{\frac{1}{2}\sum_{n, m =1} \bar N^{rs}_{nm} \a^{(r)\dag}_n \cdot
\a^{(s)\dag}_m + \sum_{n=1} \bar N^{rs}_{n0} \a^{(r)\dag}_n \cdot k^{(s)}
\Biggr\}.\label{4ssa2}
\eeq
\end{subequations}

\section{Four-String Scattering Amplitudes with the Highest Spin State}

We shall calculate the simple case first for the purpose of illustration; the SSA with three tachyons and 
the highest spin state at mass level $M^2_2 = 2(N-1)$, $N = p + q + r$ of the following form
\beq \label{highest}
\langle \Psi_1, \Psi_2, \Psi_3, \Psi_4 \vert = \langle 0, k \vert \left(e^T \cdot \a^{(2)}_1 \right)^p 
\left(e^P \cdot \a^{(2)}_1 \right)^q \left(e^L \cdot \a^{(2)}_1 \right)^r .
\eeq 
The on-shell conditions are read as 
\beq \label{onshellb}
\left(k^{(1)}\right)^2 =\left(k^{(3)}\right)^2=\left(k^{(4)}\right)^2 = 2, ~~~
\left(k^{(2)}\right)^2 = -2(N-1) . 
\eeq
This string state is generated only by primary field operators $\a_{-1}^{(2)} = \p X^{(2)}$ and 
$e^{ik^{(r)} \cdot X^{(r)}}$, $r= 1, 2, 3, 4$. If we choose the external string state as Eq. (\ref{highest}), 
the four-string scattering amplitude of the string field theory Eq. (\ref{4ssa1}) and Eq. (\ref{4ssa2}) 
reduces to 
\beq
{\cal A}_{st}^{\left(p,q,r\right)  }  &=& \int\left\vert\frac{\prod_{r=1}^4 dZ_r }{ dV_{abc}}\right\vert
\prod_{r <s} \vert Z_r -Z_s \vert^{k_r \cdot k_s + \frac{\a_s}{\a_r}\left(1-k_r^2/2\right)
+ \frac{\a_r}{\a_s}\left(1-k_s^2/2\right)} \nn\\
&& \exp \left[\sum_{r=1} \frac{\t^{(r)}}{\a_r} \left(\frac{k_r^2}{2} -1 \right) \right]
\nn \\ 
&&
\Bigl\langle 0 \Bigl\vert  \left(e^T \cdot \a^{(2)}_1 \right)^p 
\left(e^P \cdot \a^{(2)}_1 \right)^q \left(e^L \cdot \a^{(2)}_1 \right)^r \exp \left(\sum_{s\not=2} \bar N^{2s}_{10} \a^{(2)\dag}_1 \cdot k^{(s)}\right) \Bigr\vert 0 \Bigr\rangle . 
\eeq
We exclude the term $N^{22}_{10} \a^{(2)\dag}_1 \cdot k^{(2)}$ in the summation. It corresponds to the 
self-contraction which should be removed by the normal ordering. 
Decomposing the momenta $k^{(s)}$ into three polarization vectors Eq. (\ref{polarization}),
\beq
\sum_{s=1}^4 \bar N^{2s}_{10} \a^{(2)\dag}_1 \cdot k^{(s)} &=& \sum_{X = T, P, L} \left(e^X \cdot \a^{(2)\dag}_1 \right) \left(\sum_{s\not=2} \bar N^{2s}_{10} k^X_{(s)} \right),
\eeq
we have
\beq
&&\Bigl\langle 0 \Bigl\vert  \left(e^T \cdot \a^{(2)}_1 \right)^p 
\left(e^P \cdot \a^{(2)}_1 \right)^q \left(e^L \cdot \a^{(2)}_1 \right)^r \exp \left(\sum_{s\not=2} \bar N^{2s}_{10} \a^{(2)\dag}_1 \cdot k^{(s)}\right) \Bigr\vert 0 \Bigr\rangle \nn\\
&&= \left(\sum_{s\not=2} \bar N^{2s}_{10} k^T_{(s)} \right)^p \left(\sum_{s\not=2} \bar N^{2s}_{10} k^P_{(s)} \right)^q \left(\sum_{s\not=2} \bar N^{2s}_{10} k^L_{(s)} \right)^r .
\eeq 

The on-shell conditions Eq. (\ref{onshellb}) are read as  
\beq
k_1 \cdot k_2 &=& - \frac{s}{2} + N -2 , ~~~k_1 \cdot k_3 = - \frac{u}{2} -2 , ~~~
k_1 \cdot k_4 = - \frac{t}{2} -2 \nn\\
k_2 \cdot k_3 &=& - \frac{t}{2} + N -2 , ~~~ k_2 \cdot k_4 = - \frac{u}{2} + N -2,~~~
k_3 \cdot k_4 = - \frac{s}{2} -2  . 
\eeq 
It follows from the on-shell conditions that $s+t+u = 2N -8$ and 
\beq
\exp \left[\sum_{r=1} \frac{\t^{(r)}}{\a_r} \left(\frac{k_r^2}{2} -1 \right) \right]= \exp(-\t_1 N) . 
\eeq 
Using the on-shell conditions, we may write the SSA in terms of the Koba-Nielsen variable as   
\beq \label{Astkoba}
{\cal A}_{st}^{\left(p,q,r\right)  }  &=& \int_0^1 dx  \, x^{k_1 \cdot k_2} (1-x)^{k_2 \cdot k_3}\left( \frac{x}{x-1} \right)^N  \exp(-\t_1 N) \nn\\
&& \left(\sum_{s\not=2} \bar N^{2s}_{10} k^T_{(s)} \right)^p \left(\sum_{s\not=2} \bar N^{2s}_{10} k^P_{(s)} \right)^q \left(\sum_{s\not=2} \bar N^{2s}_{10} k^L_{(s)} \right)^r . 
\eeq
We only need to calculate the Neumann functions, $\bar N^{21}_{10}, ~ \bar N^{23}_{10},~ \bar N^{24}_{10}$
to get an explicit expression of the SSA in $s$-$t$ channel:
\begin{subequations} 
\beq
\bar N^{21}_{10} &=& e^{\t_1} \oint_{z=x} \frac{dz}{2\pi i} \frac{(z-1)}{z^2(z-x)} 
= e^{\t_1}\frac{x-1}{x^2} , \label{n2110} \\
\bar N^{23}_{10} &=& e^{\t_1} \oint_{z=x} \frac{dz}{2\pi i} \frac{1}{z(z-x)} = e^{\t_1}\frac{1}{x}, \label{n2310}\\
\bar N^{24}_{10} &=& e^{\t_1} \oint_{z=x} \frac{dz}{2\pi i} \frac{1}{z-Z_4} \frac{(z-1)}{z(z-x)} = 0. \label{n2410}
\eeq
\end{subequations}  
Finally, we obtain the SSA with three tachyons and the highest spin state 
\beq \label{apqr} 
{\cal A}_{st}^{\left( p,q,r\right)  }  
= \int_0^1 dx \, x^{k_1 \cdot k_2} (1-x)^{k_2 \cdot k_3} \Bigl ( \frac{k^T_1}{x} + \frac{k^T_3}{x-1} \Bigr)^p \Bigl ( \frac{k^P_1}{x} + \frac{k^P_3}{x-1} \Bigr)^q \Bigl ( \frac{k^L_1}{x} + \frac{k^L_3}{x-1} \Bigr)^r .   
\eeq
This is precisely the special case of the four-string scattering amplitude $A_{st}^{\left(  r_{n}^{T};r_{m}^{P};r_{l}^{L}\right)  }$  for three tachons and the highest spin state
which has been calculated before by using the first quantized string theory  \cite{Lai2016lauri}.

\section{Four-String Scattering Amplitudes with General Massive States}

Now we shall calculate the four-string scattering amplitudes with with three tachyons and a general massive 
state. If the external string state is chosen as 
\beq
\bigl\langle  \Psi_1, \Psi_2, \Psi_3, \Psi_4 \bigl\vert = \bigl\langle 0, k \bigr\vert  \prod_{n>0} \left(\a^{T}_n \right)^{r_n^T} 
\prod_{m>0} \left(\a^{P}_m \right)^{r_m^P} \prod_{l>0} \left(\a^{L}_l \right)^{r_l^L}
\eeq
the string field theory in the proper-time gauge brings us to the following four-string scattering amplitude
\beq
{\cal A}_{st}^{\left(  r_{n}^{T};r_{m}^{P};r_{l}^{L}\right)  } &=& \int\left\vert\frac{\prod_{r=1}^4 dZ_r }{ dV_{abc}}\right\vert
\prod_{r <s} \vert Z_r -Z_s \vert^{k_r \cdot k_s + \frac{\a_s}{\a_r}\left(1-k_r^2/2\right)
+ \frac{\a_r}{\a_s}\left(1-k_s^2/2\right)} \nn\\
&& \exp \left\{\sum_{r=1} \frac{\t^{(r)}}{\a_r} \left(\frac{k_r^2}{2} -1 \right) \right\} \Bigl\langle 0 \Bigl\vert  \prod_{n>0} \left(\a^{T}_n \right)^{r_n^T} 
\prod_{m>0} \left(\a^{P}_m \right)^{r_m^P} \prod_{l>0} \left(\a^{L}_l \right)^{r_l^L} \nn\\
&&
\exp \Biggl\{\sum_{X} \left(\a^X_{-n}\right)\left(\sum_{s\not=2} \bar N^{2s}_{n0} k^X_{(s)} \right) \Biggr\}\Bigr\vert 0 \Bigr\rangle .
\eeq
Using the on-shell conditions, we obtain 
\beq \label{Ast2nd}
{\cal A}_{st}^{\left(  r_{n}^{T};r_{m}^{P};r_{l}^{L}\right)  } &=& \int_0^1 dx \, x^{k_1 \cdot k_2} (1-x)^{k_2 \cdot k_3}\left(\frac{x}{x-1} \right)^N  \exp(-\t_1 N) \nn\\
&& \prod_{n>0} n^{r_n^T} \left(\sum_{s=1,3} \bar N^{2s}_{n0} k^T_{(s)} \right)^{r_n^T}
\prod_{m>0} n^{r_m^P}\left(\sum_{s=1,3} \bar N^{2s}_{m0} k^P_{(s)} \right)^{r_m^P}
\prod_{l>0} n^{r_l^L} \left(\sum_{s=1,3} \bar N^{2s}_{l0} k^L_{(s)} \right)^{r_l^L} .
\eeq

It is expected that the SSA of the string field theory in the proper-time gauge ${\cal A}_{st}^{\left(  r_{n}^{T};r_{m}^{P};r_{l}^{L}\right)}$ coincides with that of the first quantized theory $A_{st}^{\left(  r_{n}^{T};r_{m}^{P};r_{l}^{L}\right)}$, Eq. (\ref{Ast1st}) as in the previous case with the highest spin state.
For the purpose of comparison, let us calculate the Neumann functions 
$N^{2s}_{n0} =e^{-n\t_1} \bar N^{2s}_{n0}$, $s = 1, 3, 4$ for $n \ge 2$:
\begin{subequations}
\beq
N^{21}_{n0} &=& \frac{1}{n} \oint_{z=x} \frac{dz}{2\pi i} \frac{1}{z} 
\left(\frac{(z-1)}{z(z-x)}\right)^n \nn\\
&=& \frac{1}{n!} \left(\frac{d}{dz}\right)^{n-1} \left(\frac{(z-1)^n}{z^{n+1}}\right) \Biggl\vert_{z=x}, \\ 
N^{23}_{n0} &=& \frac{1}{n}  \oint_{z=x} \frac{dz}{2\pi i} \frac{1}{z-1} 
\left(\frac{(z-1)}{z(z-x)}\right)^n \nn\\
&=& \frac{1}{n!} \left(\frac{d}{dz}\right)^{n-1} \left(\frac{(z-1)^{n-1}}{z^n}\right) \Biggl\vert_{z=x}, \\
N^{24}_{n0} &=& \oint_{z=x} \frac{dz}{2\pi i} \frac{1}{z-Z_4} 
\left(\frac{(z-1)}{z(z-x)}\right)^n  \nn\\
&=& 0.
\eeq 
\end{subequations}
For $n=2$, we have 
\beq \label{neumannn=2}
N^{21}_{20} &=& \frac{1}{2!} \frac{d}{dz} \frac{(z-1)^2}{z^3} \Bigl\vert_{z=x} = - \frac{1}{2} \left\{
\frac{(1-x)^2}{x^4} + \frac{2(1-x)}{x^4} \right\} \nn\\
N^{23}_{20} &=& \frac{1}{2!} \frac{d}{dz} \frac{(z-1)}{z^2} \Bigl\vert_{z=x} = -\frac{1}{2} \left\{
\frac{1}{x^2} - \frac{2}{x^3} \right\} , \\
N^{24}_{20} &=& 0 . \nn
\eeq 
Comparing the SSA of the string field theory ${\cal A}_{st}^{\left(r_{n}^{T};r_{m}^{P};r_{l}^{L}\right)}$ Eq. (\ref{Ast2nd}) with Eq. (\ref{neumannn=2}) with that of the first quantized theory 
Eq. (\ref{Ast1st}), we immediately realize that they do not agree with 
each other in contrast to our expectations. 

We may recall that the SSA of the string field theory is defined on the string world sheet whereas the 
SSA of the first quantized theory is defined on the upper half plane. For the string states corresponding to the primary operators, the SSA of the string field theory is constructed to be invariant under the 
conformal transformation. This was confirmed in the last section by comparing the SSA of the 
string field theory for the highest spin state, ${\cal A}_{st}^{\left( p,q,r\right)}$ with that of the 
SSA of the first quantized theory. However, we must pay attention to 
\beq
\a_{-n}^{(r)} \vert 0 \rangle = \lim_{z \rightarrow Z_r} \frac{\p^n X}{(n-1)!} \vert 0 \rangle, ~~~
\text{for}~~ n \ge 2, 
\eeq 
being a non-primary operator. If the external string states are generated by the non-primary field operators 
such as the higher derivatives of the scalar fields, the SSA of the string field theory may not be invariant 
under the conformal transformation. Thus, in the case of the SSA with the general massive state, 
we should carefully take the conformal transformation of the non-primary operators into account.

Let us re-examine the equivalence of the SAA of the string field theory ${\cal A}_{st}^{\left( p,q,r\right)}$ and that of the first quantized theory for the highest spin state, $A_{st}^{\left( p,q,r\right)}$, paying 
attention to the conformal transformation between the world sheet ${\mathbb W}$ and the upper half plane ${\mathbb H}$.   
We may rewrite the SSA $A_{st}^{\left( p,q,r\right)}$ which defined on the upper half plane ${\mathbb H}$ as 
\beq \label{AstH}
A_{st}^{\left( p,q,r\right)} &=& {\cal A}_{st}^{\left( p,q,r\right)}\vert_{\mathbb H} \nn\\
&=& \int_0^1 dx \, x^{k_1 \cdot k_2} (1-x)^{k_2 \cdot k_3} \Bigl ( \sum_{s=1,3} \bar N^{2s}_{10}\vert_{\mathbb H} k^T_{(s)} \Bigr)^p \Bigl ( \sum_{s=1,3} \bar N^{2s}_{10}\vert_{\mathbb H} k^P_{(s)} \Bigr)^q \Bigl ( \sum_{s=1,3} \bar N^{2s}_{10}\vert_{\mathbb H} k^L_{(s)} \Bigr)^r    
\eeq
where we define the Neumann functions on the upper half plane $\bar N^{rs}_{10} \vert_{\mathbb H}$ as 
\beq
\bar N^{21}_{10} \vert_{\mathbb H} = \frac{1}{x}, ~~~ \bar N^{23}_{10} \vert_{\mathbb H} = \frac{1}{x-1} . 
\eeq 
Under the conformal transformation from the second string patch to the upper half plane
\beq \label{conformal2nd}
\o &=& e^{\zeta_2} = e^{-\t_1} \frac{z(z-x)}{(z-1)} = e^{-\t_1} \left\{ z + 1 -x + \frac{1-x}{z-1} \right\},
\eeq
the primary field operator $\p X$ transforms as
\beq
\frac{\p X}{\p z} &=& \frac{\p \o}{\p z} \frac{\p X}{\p \o} = e^{-\t_1} \left(\frac{x}{x-1}\right) \frac{\p X}{\p \o}.
\eeq   
Because the Neumann function corresponds to a Fourier component of the Green's function of the scalar field,
the Neumann function $\bar N^{rs}_{10}$ also transforms as $\p X$ under the conformal transformation: 
\beq
\bar N^{2s}_{10} \vert_{\mathbb H}  &=& e^{-\t_1} \left(\frac{x}{x-1}\right) \bar N^{2s}_{10} \vert_{\mathbb W}, ~~~
s = 1, 2, 3, 4, 
\eeq
where $\bar N^{2s}_{10} \vert_{\mathbb W}$ denotes the Neumann function on the world sheet of strings Eq. (\ref{n2110}) and Eq. (\ref{n2310})
\beq \label{n2110w}
\bar N^{21}_{10} \vert_{\mathbb W} = e^{\t_1} \left(\frac{x-1}{x^2}\right), ~~~
\bar N^{23}_{10} \vert_{\mathbb W} = e^{\t_1} \frac{1}{x} . 
\eeq
Rewriting the SAA on the upper half plane ${\cal A}_{st}^{\left( p,q,r\right)}\vert_{\mathbb H}$ Eq. (\ref{AstH}) in  
terms of the Neumann function on the world sheet $\bar N^{2s}_{10} \vert_{\mathbb W}$ Eq. (\ref{n2110w}), we 
find that it can be identified as the SAA on the world sheet of the string field theory 
${\cal A}_{st}^{\left( p,q,r\right)}\vert_{\mathbb W}$ Eq. (\ref{Astkoba})
\beq
{\cal A}_{st}^{\left( p,q,r\right)}\vert_{\mathbb H} &=& \int_0^1 dx \, x^{k_1 \cdot k_2} (1-x)^{k_2 \cdot k_3}
\left(\frac{x}{x-1}\right)^N \exp\left(-N\t_1\right) \nn\\
&&\Bigl ( \sum_{s=1,3} \bar N^{2s}_{10}\vert_{\mathbb W} k^T_{(s)} \Bigr)^p \Bigl ( \sum_{s=1,3} \bar N^{2s}_{10}\vert_{\mathbb W} k^P_{(s)} \Bigr)^q \Bigl ( \sum_{s=1,3} \bar N^{2s}_{10}\vert_{\mathbb W} k^L_{(s)} \Bigr)^r \nn\\
&=& {\cal A}_{st}^{\left( p,q,r\right)}\vert_{\mathbb W}
\eeq 

The four-string scattering amplitude with three tachyons and a general string state in the $s$-$t$ channel
$A_{st}^{\left(r_{n}^{T};r_{m}^{P};r_{l}^{L}\right)}$ Eq. (\ref{Ast1st}) may be also rewritten 
as follows 
\beq
A_{st}^{\left(  r_{n}^{T};r_{m}^{P};r_{l}^{L}\right)} &=& {\cal A}_{st}^{\left(  r_{n}^{T};r_{m}^{P};r_{l}^{L}\right)}\vert_{\mathbb H} \nn\\
&=& \int_0^1 dx \, x^{k_1 \cdot k_2} (1-x)^{k_2 \cdot k_3} \nn\\
&& \prod_{n>0} \left(\sum_{s=1,3} N^{2s}_{n0}\vert_{\mathbb H} k^T_{(s)} \right)^{r_n^T}
\prod_{m>0} \left(\sum_{s=1,3} N^{2s}_{m0}\vert_{\mathbb H} k^P_{(s)} \right)^{r_m^P}
\prod_{l>0} \left(\sum_{s=1,3} N^{2s}_{l0}\vert_{\mathbb H} k^L_{(s)} \right)^{r_l^L}
\eeq 
we define the Neumann functions on the upper half plane $N^{2s}_{m0}\vert_{\mathbb H}$, $s = 1, 3, 4$:
\beq \label{n21upper}
N^{21}_{n0}\vert_{\mathbb H} &=& \frac{(-1)^{n-1}(n-1)!}{x^n}, ~~~ N^{23}_{n0}\vert_{\mathbb H} = 
\frac{(-1)^{n-1}(n-1)!}{(x-1)^n}, ~~~ N^{24}_{n0}\vert_{\mathbb H}= 0.
\eeq  
Now we shall examine the transformation properties of the Neumann functions $\bar N^{21}_{n0}$, $n \ge 2$ under the 
conformal mapping. 
Let us consider the case of $\bar N^{21}_{20}\vert_{\mathbb H}$ first. Because the Neummann function $\bar N^{21}_{20}\vert_{\mathbb H}$ is the 2nd Fourier component of the Green's function of the scalar field, it 
transforms as the non-primary operator $\frac{1}{2!}\p^2 X$ under the conformal transformation: From the transformation 
of $\p^2 X$ under the conformal transformation from the world sheet to the upper half plane
\beq
\frac{\p^2 X}{\p z^2} &=& \left(\frac{\p \o}{\p z} \right)^2 \frac{\p^2 X}{\p \o^2} + \left(\frac{\p \o}{\p z}\right) \frac{\p}{\p \o} \left(\frac{\p \o}{\p z} \right) \left(\frac{\p X}{\p \o} \right) \nn\\
&=& e^{-2\t_1} \left(\frac{x}{1-x}\right)^2 \frac{\p^2 X}{\p \o^2} - e^{-\t_1} \frac{2}{(1-x)^2} \left(\frac{\p X}{\p \o} \right), 
\eeq 
it follows that the $\bar N^{2s}_{20}\vert_{\mathbb H}$ is related to the Neumann function on the world sheet 
$\bar N^{2s}_{20}\vert_{\mathbb W}$ as 
\beq
2! \bar N^{2s}_{20}\vert_{\mathbb H} &=& e^{-2\t_1} \left(\frac{x}{1-x}\right)^2 2! \bar N^{2s}_{20}\vert_{\mathbb W} 
- e^{-\t_1} \frac{2}{(1-x)^2} \bar N^{2s}_{10}\vert_{\mathbb W}. 
\eeq
A simple algebra brings us  
\begin{subequations}
\beq
2! \bar N^{21}_{20}\vert_{\mathbb H} &=& e^{-2\t_1} \left(\frac{x}{1-x}\right)^2 2 (-1) \frac{e^{2\t_1}}{2} \left\{ \frac{(1-x)^2}{x^4} + 
\frac{2(1-x)}{x^4} \right\} + e^{-\t_1} \left(- \frac{2}{(1-x)^2}\right) e^{\t_1} \left(\frac{x-1}{x^2}\right) \nn\\
&=& - \frac{1}{x^2}, \label{simplea}\\
2! \bar N^{23}_{20} \vert_{\mathbb H}
&=&  e^{-2\t_1} \left(\frac{x}{1-x}\right)^2 2 \frac{e^{2\t_1}}{2} \left\{ - \frac{1}{x^2} + \frac{2}{x^3} \right\} +  e^{-\t_1} \left(- \frac{2}{(1-x)^2}\right) e^{\t_1} \frac{1}{x} \nn\\
&=& - \frac{1}{(1-x)^2}. \label{simpleb}
\eeq 
\end{subequations} 
By comparing Eqs. (\ref{simplea},\ref{simpleb}) and Eq. (\ref{n21upper}), we find that the Neumann functions on the upper half plane
are related to the Neumann functions on the world sheet by the conformal transformation: 
\beq
N^{21}_{20}\vert_{\mathbb H} = 2! \bar N^{21}_{20}\vert_{\mathbb H}, ~~~
N^{23}_{20}\vert_{\mathbb H} = 2! \bar N^{23}_{20}\vert_{\mathbb H}.
\eeq 
The factor $2!$ may be due to different notational conventions. 

We may check the relations between the Neumann functions of two string theories for $n=3$ by using the 
explicit form of the conformal transformation of $\p^3 X$,
\beq
\frac{\p^3 X}{\p z^3} &=& \left(\frac{\p w}{\p z}\right)^3 \frac{\p^3 X}{\p w^3} + 3 \left(\frac{\p^2 w}{\p z^2}\right)\left(\frac{\p w}{\p z}\right) \frac{\p^2 X}{\p w^2} + 
\left(\frac{\p^3 w}{\p z^3}\right)\frac{\p X}{\p w}.
\eeq 
Because the Neumann function $\bar N^{2s}_{30}\vert_{\mathbb H}$ transforms as $\frac{1}{3!} \p^3 X$:
\beq
3! \bar N^{2s}_{30}\vert_{\mathbb H} &=& \left(\frac{\p w}{\p z}\right)^3 3! \bar N^{2s}_{30}\vert_{\mathbb W} + 3 \left(\frac{\p^2 w}{\p z^2}\right)\left(\frac{\p w}{\p z}\right) 2! \bar N^{2s}_{20}\vert_{\mathbb W} + 
\left(\frac{\p^3 w}{\p z^3}\right) \bar N^{s1}_{10}\vert_{\mathbb W}, ~~~ s =1, 3.
\eeq 
By an explicit calculation we have 
\beq \label{n2sexplicit}
3!\, \bar N^{2s}_{30}\vert_{\mathbb H} &=& \left(  \frac{e^{-\tau_{1}}x}{x-1}\right)  ^{3}\cdot\left\{  \frac
{2e^{3\tau_{1}}\left(  x-1\right)  \left(  x^{4}-8x^{3}+\left(  9Z_{s}%
+10\right)  x^{2}-\left(  3Z_{s}+15\right)  Z_{s}x+6Z_{s}^{2}\right)  }%
{x^{5}\left(  x-Z_{s}\right)  ^{3}}\right\}  \nonumber\\
&&  +3\frac{e^{-\tau_{1}}x}{x-1}\frac{\left(-2e^{-\tau_{1}}\right)}{\left(  x-1\right)
^{2}}\cdot\left\{-\frac{e^{2\tau_{1}}\left(  x-1\right)  \left(
x^{2}-3x+2Z_{s}\right)  }{x^{3}\left(  x-Z_{s}\right)  ^{2}}\right\}
+\frac{6e^{-\tau_{1}}x}{\left(  x-1\right)  ^{3}}\frac{e^{\tau_{1}}\left(
x-1\right)  }{x\left(  x-Z_{s}\right)  }\nonumber\\
&=& \frac{2}{\left(  x-Z_{s}\right)  ^{3}} . 
\eeq 
Comparing Eq. (\ref{n2sexplicit}) with Eq. (\ref{n21upper}) leads us to
\beq  
N^{2s}_{30} \vert_{\mathbb H} = 3! \bar N^{2s}_{30} \vert_{\mathbb H}, ~~~ s = 1, 3. 
\eeq 

These algebras to compare the Neumann functions $N^{2s}_{n0}\vert_{\mathbb H}$ of the first quantized theory 
and the Neumann functions $\bar N^{2s}_{n0}\vert_{\mathbb W}$ for $n=2, 3$ strongly suggest us that 
in general for $n \ge 1$, 
\beq
N^{2s}_{n0} \vert_{\mathbb H} = n! \bar N^{2s}_{n0} \vert_{\mathbb H}, ~~~ s = 1, 3. 
\eeq 
It follows from the Fa\'{a} di Bruno's formula \cite{Johnson} that the derivative of the scalar field operator $\p^n X$, $n \ge 2$ 
transforms under the conformal transformation Eq. (\ref{conformal2nd}) as follows:
\beq 
\frac{\partial^{n}}{\partial z^{n}}X\left(  \omega\left(  z\right)  \right)
&=& n!\sum_{k=1}^{n}\frac{\partial^{k}X}{\partial\omega^{k}}\sum_{\left\{
k_{m}\right\}  }\left\{  \frac{1}{k_{1}!\cdots k_{n}!}\prod_{m=1}^{n}\left(
\frac{1}{m!} \frac{d^m w}{d z^m}\right)  ^{k_{m}}\right\} \\
&&
\text{ with }\sum_{m=1}^{n}mk_{m}=n\text{ and }\sum_{m=1}^{n}k_{m}=k . \nn
\eeq
From it we deduce that the Neumann function $N^{2s}_{n0}\vert_{\mathbb H}$ transforms as 
\beq
n! \,N^{2s}_{n0}\vert_{\mathbb H} &=& n!\sum_{k=1}^{n}k! \, \bar{N}%
_{k0}^{2s}\vert_{\mathbb W} \sum_{\left\{  k_{m}\right\}  }\left\{  \frac{1}{k_{1}!\cdots k_{n}%
!}\prod_{m=1}^{n}\left( \frac{1}{m!} \frac{d^m w}{d z^m}\right)
^{k_{m}}\right\}.
\eeq 
Rewriting $\bar{N}_{k0}^{2s}\vert_{\mathbb W}$ as 
\beq
k! \, \bar{N}_{k0}^{2s}\vert_{\mathbb W} 
& =&\frac{\left(  k-1\right)  !}{2\pi i}\oint_{z=x}\frac{1}{\omega^{k}\left(
z\right)  }d\left(  \ln\left\vert z-Z_{s}\right\vert \right) \nn \\
& =&-\frac{k!}{2\pi i}\oint_{\omega=0}\frac{\ln\left\vert z\left(
\omega\right)  -Z_{s}\right\vert }{\omega^{k+1}}d\omega \nn\\
& =&-\left.  \frac{d^{k}}{d\omega^{k}}  \ln\left\vert z\left(
\omega\right)  -Z_{s}\right\vert  \right\vert _{\omega=0},
\eeq 
we have 
\beq
n! \,N^{2s}_{n0}\vert_{\mathbb H} &=& -n!\sum_{k=1}^{n}\left.  \frac{d^{k}%
}{d\omega^{k}}\ln\left\vert z\left(  \omega\right)  -Z_{s}\right\vert
\right\vert _{\omega=0}\sum_{\left\{  k_{m}\right\}  }\left[
\frac{1}{k_{1}!\cdots k_{n}!}\prod_{m=1}^{n}\left( \frac{1}{m!} \frac{d^m w}{d z^m}\right)
^{k_{m}}\right] \nn \\
&=&-\left.
\frac{d^{n}}{dz^{n}}\left(  \ln\left\vert z-Z_{s}\right\vert \right)
\right\vert _{z=x} \nn\\
&=& \frac{\left(  -1\right)  ^{n-1}\left(  n-1\right)
!}{\left(  x-Z_{s}\right)  ^{n}} \nn\\
&=& \bar N^{2s}_{n0} \vert_{\mathbb H}. 
\eeq
This proves the relationship between the Neumann functions of the first quantized theory and those of the 
second quantized theory. The SSA of the deformed cubic string field theory is related to the 
SSA of the conventional first quantized string theory by the conformal transformation which maps
the string world sheet to the upper half plane.

\section{Conclusions}

We applied the deformed cubic bosonic open string field theory to the four-string scattering amplitudes with 
three tachyons and one general string state, which are the main objects of the recent studies on the 
high-energy symmetries of string theory. String scattering amplitudes have been calculated in the previous works mostly by using the first quantized string theory. The reason may be that little is known about
the Fock space representations of the multiple string vertices of the covariant cubic open string field 
theory. The Fock space representation of the four-string vertex has been constructed before by Giddings
in Ref. \cite{Giddings86} by mapping the world sheet onto the upper half plane. However, this transformation 
function by which the non-planar world sheet is mapped onto the upper half plane, 
has a non-trivial structure such as branch cuts. It is not practical to use this Fock space representation of the four-string vertex to calculate the string scattering amplitudes due to its complexity.
It is also unknown how to generalize this construction to multiple string vertices of more than four strings.
One of the authors has proposed the deformed cubic open string field theory \cite{TLee2017plb} as a proper remedy to these drawbacks of previous approaches. If we deform the external string states while 
the cubic string field action is kept intact, the world sheet of string scattering can be made planar.
This consistent deformation greatly helps us to construct the Fock space representations of the multiple
string vertices systematically by making use of the light-cone string field theory technique which is 
readily available. The deformed cubic string field theory has been adopted to calculate the zero-slope 
limit of the covariant string field theory on multiple space filling $D$-branes: 
In the zero-slope limit, the deformed cubic string field theory is shown to reduce to the non-Abelian Yang-Mills theory correctly. It is noteworthy that this result is exact and obtained without using the level truncations or the field redefinitions. The zero-slope limit of the deformed cubic string field theory
on multiple $Dp$-branes was also calculated. The deformed cubic string field theory on multiple $Dp$-branes 
is shown to reduce to the $U(N)$ matrix-valued scalar field theory interacting with $U(N)$ Yang-Mills gauge fields in the zero-slope limit.  

In the present work, we calculated the string scattering amplitudes by using the deformed cubic string 
field theory and compared the obtained results with those of previous works based on the first quantized string theory  \cite{Lai2016lauri}. If we choose the highest spin state as one of the external string states, the string 
scattering amplitude calculated by using the deformed cubic string field theory is in perfect agreement 
with the results obtained by the first quantized theory \cite{Lai2016lauri}. 
However, for the general massive string 
state, the deformed string field theory yields a scattering amplitude which differs from that of the 
first quantized string theory. The origin of this discrepancy is found to be the transformtion of the Neumann functions under the conformal mapping from the world sheet onto the upper half plane. 
The cubic string field theory is initially defined on the world sheet of string scattering whereas the conventional SSA of the first quantized string theory is defined on the upper half plane. For the general massive string states, we need to take into account of the non-trivial behavior of the Neumann functions 
under the conformal transforamtion. The Neumann functions are Fourier components of the Green's function for the scalar field: The $n$-th component $\bar N^{rs}_{n0}$ 
transforms like $\p^n X$ under the conformal mapping. For the highest spin states which are generated only by 
the primary field operator $\p X$, the Neumann function $\bar N^{rs}_{10}$ transforms like a primary field 
operator. In this case, the scattering amplitudes of both string theories, being invariant under the conformal 
transformation, agree with each other. However, for the massive string states, which are generated by 
non-primary operators $\p^n X$, $n \ge 2$, the corresponding Neumann functions $\bar N^{rs}_{n0}$
transform like non-primary field operator $\p^n X/n!$, $n \ge 2$. By an explicit calculation, we confirmed 
this and showed that the scattering amplitudes of the deformed cubic string field theory are 
releated to those of the first quantized string theory by the conformal transformation from the world sheet onto the upper half plane for the cases with $n=2$ and $n=3$. By some algebra, we proved that this 
relationship between the scattering amplitude of the second quantized theory and that of the first quantized theory holds for general massive string states. This work shows how one can make use of the covariant cubic (deformed) string field theory to calculate the scattering amplitudes. It also enables one to convert the scattering amplitudes evaluated on the upper half plane to those on the string world sheet for the string states generated by the non-primary operators. 

We may extend the present study on the string scattering amplitude with three tachyons and one tensor state to more general string scattering amplitudes with multi-tensors.   
In this case, we may encounter normal odered products of derivatives of the scalar field operators such as 
$:\p^n X \p^m X:$, $n, m \ge 1$ which transform non-trivially under the conformal transformation thanks to the conformal anomaly \cite{Polyakov2017}.

\vskip 1cm

\begin{acknowledgments}
This work was supported in part by the Ministry of Science and Technology and
S. T. Yau center of NCTU, Taiwan.
TL was supported by Basic Science Research Program through the National Research Foundation of Korea(NRF) funded by the Ministry of Education (2017R1D1A1A02017805) and partly by 2017 Research Grant from Kangwon National University (No. D1001513-01-01). 
\end{acknowledgments}


%

%





\end{document}